\documentstyle[12pt]{article}
\newcommand{\NP}[1]{Nucl. \ Phys.}
\newcommand{\PL}[1]{Phys. \ Lett.}
\newcommand{\p}[1]{\partial}
\newcommand{\PRL}[1]{Phys.\ Rev.\ Lett. }
\newcommand{\MPL}[1] { Mod. Phys. Lett. }
\newcommand{\IJMP}[1] { Int. J. Mod. Phys. }
\begin{document}
\title{
\begin{flushright}
{\small SMI-01-95 \\
HEP-TH/9502000}
\end{flushright}
\vspace{2cm}
 $N$-point amplitudes for d=2 c=1 Discrete States \\
from String Field Theory.
}
\author{
I.Ya.Aref'eva, \thanks{Steklov Mathematical Institute,
Vavilov 42, GSP-1, 117966, Moscow, e-mail: Arefeva@qft.mian.su} \\
P.B.Medvedev \thanks{
 Institute of Theoretical and Experimental Physics,117259, Moscow }\\ and\\
 A.P.Zubarev \thanks{Steklov Mathematical Institute,
Vavilov 42, GSP-1, 117966, Moscow, e-mail: Zubarev@qft.mian.su}}
\maketitle
\begin{abstract}
Starting from string field theory for 2d gravity coupled to c=1 matter
we analyze N-point off-shell tree amplitudes of discrete states. The
amplitudes exhibit the pole structure and we use the oscillator
representation to extract the residues.  The residues are generated by
a simple effective action. We show that the effective action can be
directly deduced from a string field action in a special
transversal-like gauge.
\end{abstract}

\section{Introduction}
In this paper we continue the study of the scattering amplitudes of discrete
states in
2d string that was started in our previous paper \cite{amz3}, which in the
following will be refereed as {\bf I}.  The characteristic  property of
 the model is an appearance of discrete states  which together with
tachyon states exhaust the spectrum of physical states
\cite{Gross,Pol,Lian,az1}.  Our starting  point in {\bf I} is the Witten-type
String Field Theory \cite{Witten} for this model \cite{az1}. The main reason
to employ String Field Theory for this model is a necessity to have a
definition for off-shell scattering amplitudes since the on-shell amplitudes
are ill-defined \cite{az2}. The divergences of the amplitudes are caused by the
resonance denominator phenomenon well-known in the usual field theories
and occurred in the case of particle decays. A simple example is the Lee
model \cite{sh}.  Indeed, two-vectors of momentum-energy of discrete states
form a
two-dimensional lattice  and summing up a pair of discrete momenta
(according to the conservation law with the background charge) one gets as
a rule the point of this lattice again. In other words, the internal lines
of the graphs for tree-level amplitudes describe a propagation of real
(non-virtual) "particles".  Hence, calculating a scattering amplitude one
unavoidably encounters the poles that leads to  senseless results.

 Starting with String Field Theory  we have defined in {\bf I}
the 4-point scattering amplitudes off-shell
and then have investigated their behavior near the mass-shell.
In this paper we consider the N-point off-shell scattering
amplitudes for discrete states and an effective action for the residues.
In I we have used intensively  the off-shell
conformal method developed by Samuel et al. \cite{sam}. However,
application of this method to the case of the $N$-point scattering
amplitude meets  serious analytical difficulties. By this reason here we
will use the oscillator representation of 2d String Field Theory
\cite{pt,kaku}.  It turns out that it is possible to separate the most
singular part of N-point amplitudes.  Moreover, the most singular part is
determined by the structure constant of the discrete states operator
algebra, which as it was shown in \cite{KP}, is responsible for the hidden
symmetry of the model. This gives a hint that the {\em most singular
parts} of scattering amplitudes can be described by a simple effective
theory.  Note in this context that a few years ago Klebanov and Polyakov
\cite{KP} proposed an effective action to describe the interaction of
discrete states. To search for some sort of effective action we use a new
gauge which we call a transversal-like gauge.  We find that the most
singular parts of tree-level discrete states off-shell amplitudes are
generated by the part of the gauge-fixed action containing the discrete
states fields only. We identify this piece of the gauge-fixed action with
an effective action of the theory.

The paper is organized as follows. In Section 2 we investigate the
singular behavior of an arbitrary N-point tree-level off-shell discrete
states amplitude near mass-shell by using the oscillator representation of
String Field Theory in the Siegel gauge.  We find that the most singular
part of the amplitude is the product of the poles in chanel variables
times the product of structure constants, which enters in OPE of external
states vertex operators.  In Section 3 we investigate the 2d String Field
Theory in the new gauge and separate the part responsible for the most
singular parts of the graphs.
%%%%%%%%%%%%%%%%%%%%%%%%%%%%%%%%%%%%%%%%%%%%%%%%%%%%%%%%%%%%%
\section{The $N$-point tree-level off-shell discrete states amplitudes
in the Siegel gauge}
In this section we shall describe the tree graphs of the model
by using the oscillator representation  of 2d String Field Theory.

The Feynman graphs we are going to examine are generated by the
Witten-type action \cite{Witten}
\begin{equation}
\label{s1}
S=\int (\Phi \ast Q \Phi +
\frac{2}{3}\Phi \ast \Phi \ast \Phi )
\end{equation}
As usual, $\Phi$ is a string field and $\int$ and $\ast$ are well-
known Witten integration and product operations. The action can be
written in alternative form
\begin{equation}
\label{s2}
S=\, _1\langle \,_2\langle I|
|\Phi \rangle _1 Q^{(2)} |\Phi\rangle
_2 +
\frac{2}{3}\,_1\langle \,_2\langle \,_3\langle V|
|\Phi \rangle_1  |\Phi\rangle_2 |\Phi \rangle_3,
\end{equation}
where
$\, _1\langle \,_2\langle I|$
is "identity" operator transforming kets into  bras and
$\,_1\langle \,_2\langle \,_3\langle V|$
is Witten three-vertex describing the interaction of string.
The explicit form and properties for
$\, _1\langle \,_2\langle I|$
and
$\,_1\langle \,_2\langle \,_3\langle V|$
for $D=2$ had been discussed in \cite{pt,kaku}.
The propagator in Siegel gauge  $b_0 |\Phi\rangle =0$  is equal to
$b_0(L_0)^{-1}$.

In the modern setting, the discrete states appear as a non-trivial
cohomology of BRST charge $Q$ : $Q\vert \psi\rangle=0 $,
$\vert \psi\rangle \ne Q\vert \lambda\rangle $, and they can be classified
according to their ghost number. The nontrivial cohomology can be found
for ghost number $0,...,3$ and it can be proved that not only the states with
fixed momenta are in the spectrum but the whole spectrum is exhausted by
these states plus the tachyon with non-discrete momentum.
In the following we shall deal with compactified version of the
model with matter coordinate $x$ being compactified on $SU(2)$
radius. In this case all points of the spectrum can be described in
unified fashion and belong to a single lattice.
This physical states in the total ("matter" + ghosts) Fock space can be
explicitly described in two ways: by using Shur polynomials \cite{bou} or
in so called "material gauge" in terms of $SU(2)$ raising and lowering
operators \cite{KP,w,wz}. For our purposes the second description is more
convenient.

The on-shell physical states are defined in terms of
conformal fields $Y$  \cite{KP,w}
\begin {equation}
\label {14}
Y^{\pm}_{J,n}(z)=cW^{\pm}_{J,n}(z)=cV_{J,n}e^{\sqrt{2}(1\mp J)}.
\end   {equation}
The momenta of discrete states read:
\begin{equation}
(p,\varepsilon)=\sqrt{2}(n,~1\mp J)
\label{9}
\end{equation}
where $$J=0,\frac{1}{2},1,\frac{3}{2},2,...;~~n=-J,-J+1,...,J.$$
Hereafter we adopt the standard correspondence between conformal
fields $Y_{J,n}^{\sigma}(z)$ $ ( \sigma = \pm ) $ and states
$|Y_{J,n}^{\sigma}\rangle$ in the total Fock space:
\begin{equation}
\lim
_{z\to 0}Y_{J,n}^\sigma (z)|0\rangle
=|Y_{J,n}^\sigma\rangle ,
\label{cor}
\end{equation}
where $|0\rangle$ is $SL_2$-invariant  vacuum.

Hidden symmetry of the theory is encoded in structure constants of
OPE for $W$-s:
\begin{equation}
\label{ope}
W^{\sigma _1}_{J_1,n_1}(z)W^{\sigma _2}_{J_2,n_2}(0)
=\frac{1}{z}
{\tilde f}^{J_3 n_3 \sigma _3}_{J_1 n_1 \sigma _1;~J_2 n_2
\sigma _2}
W^{\sigma_3}_{J_3,n_3}(0)~+~...~.
\end{equation}
The explicit expressions for $f$-s had been obtained in \cite{KP}:
$$
{\tilde f}^{J_3 n_3 +}_{J_1 n_1 +;~J_2 n_2
+}=(J_2n_1-J_1n_2)\delta_{J_1+J_2-1,J_3}
\delta_{n_1+n_2,n_3},
$$
\begin{equation}
\label{ff}
{\tilde f}^{J_3 n_3 -}_{J_1 n_1 +;~J_2 n_2 -}
=(-J_1n_3-J_3n_1)\delta _{J_1+J_3-1,J_2}\delta_{n_3-n_1,n_2}
\end{equation}
and other $f$-s are equal to zero. It has been found in \cite{ms} that
naive OPE of the vertex operators $W$ does form well defined current
algebra. To avoid this one need to multiply each of $W$ by corresponding
cocycle operator.

However the String Field Theory provides a rigid frame for string field
and we cannot multiply by hand the vertex operators, which are on-shell
string fields, by some cocycles. Then the expression for the structure
constant in OPE of vertex operators are modified to be
\begin{equation}
\label{fff}
 f^{J_3 n_3 \sigma _3}_{J_1 n_1 \sigma _1;~J_2 n_2 \sigma _2}
=(-1)^{2J_1(J_2-n_2-1)}
{\tilde f}^{J_3 n_3 \sigma _3}_{J_1 n_1 \sigma _1;~J_2 n_2 \sigma _2}.
\end{equation}
The structure constants (\ref{fff}) does not poses definite
symmetry properties of indices, therefore they are not correspond
to any algebra.

To set the convention we take any graph in the form:
$$
A=\,
_1\langle \, _2 \langle \, _{i_1}\langle V |
\,\,\, _{j_1}\langle \, _3 \langle \, _{i_2}\langle V |\, \dots \,\,\,
_{j_{N-3}}\langle \, _{N-1} \langle  \, _{N}\langle V |
$$
\begin{equation}
\frac{b^{j_1}_0}{L^{j_1}_0}
|I\rangle _{i_1} \rangle _{j_1}\,\,\,
\frac{b^{j_2}_0}{L^{j_2}_0}|I\rangle _{i_2} \rangle _{j_2}\, \dots \,
\,\,\frac{b^{j_{N-3}}_0}{L^{j_{N-3}}_0}|I\rangle _{i_{N-3}}
\rangle _{j_{N-3}}\,\,\,
|Y _1\rangle _1 \,\,\, |Y_2\rangle _2\,\dots \,\,\,
|Y_N\rangle _N
\label{gr}
\end{equation}

In the Liouville zero modes sector of Fock space we put the following
normalization
\begin{equation}
\langle \varepsilon '| \varepsilon \rangle
=\delta (\varepsilon '- \varepsilon )
\label{norm}
\end{equation}

\begin{equation}
p_{\phi}|\varepsilon \rangle =\varepsilon | \varepsilon \rangle
\label{state}
\end{equation}

To describe the proper variables for off-shell tree amplitudes we specify
the way of moving off the mass shell for discrete states. Although we are not
going
to use the methods of Conformal Field Theory we adopt the same definition for
off-shell discrete states as in our previous paper \cite{amz3} taking the
discrete states in the form \begin{equation}
Y^{\pm}_{J,n}=cV_{J,n}e^{\sqrt{2}(1\mp J)\phi} \label{17} \end{equation}
we can relax the mass-shell condition by simply substituting:
$\sqrt{2}(1\mp J)\phi  \to \varepsilon \phi$ for some parameter
$\varepsilon $. The fields $$Y^{\varepsilon}_{J,n,\sigma}=
cV_{J,n}e^{\varepsilon \phi}
$$
will be associated with external legs
of tree graphs. The subscript $\sigma =\pm$ indicates that $\varepsilon$
lies in a small neighborhood of the point $\sqrt{2}(1-\sigma J)$.
It is obvious that as soon as the couple of external states
$Y^{\varepsilon}_{J,n,\sigma}$
is given, the energy-momenta of any internal line is uniquely defined.
In the following we shall use the compact notation for subscripts of
discrete states $(J,n,\sigma )=a$.  For an amplitude of a tree graph with
$N$ external off-shell states $Y^{\varepsilon_r}_{a_r}$, $~r=1,\dots ,N$
we use the notation $A_N(a_1,...,a_N)$.

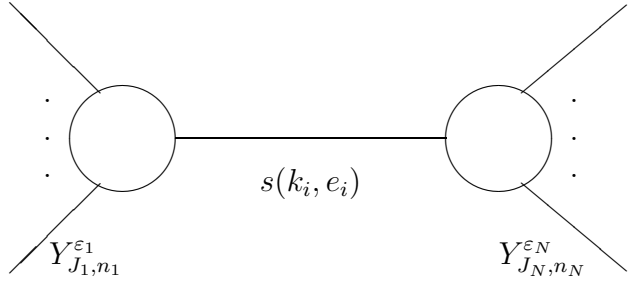
\begin{figure}
\begin{center}
\unitlength=1.0mm
\special{em:linewidth 0.4pt}
\linethickness{0.4pt}
\begin{picture}(117.00,48.00)
\put(50.00,30.00){\circle{14.00}}
\put(100.00,30.00){\circle{14.00}}
\put(47.00,36.00){\line(-1,1){12.00}}
\put(47.00,24.00){\line(-1,-1){12.00}}
\put(103.00,36.00){\line(6,5){14.00}}
\put(103.00,24.00){\line(6,-5){14.00}}
\put(93.00,30.00){\line(-1,0){36.00}}
\put(40.00,35.00){\makebox(0,0)[cc]{.}}
\put(40.00,30.00){\makebox(0,0)[cc]{.}}
\put(40.00,25.00){\makebox(0,0)[cc]{.}}
\put(110.00,35.00){\makebox(0,0)[cc]{.}}
\put(110.00,30.00){\makebox(0,0)[cc]{.}}
\put(110.00,25.00){\makebox(0,0)[cc]{.}}
\put(75.00,24.00){\makebox(0,0)[cc]{$s(k_i,e_i)$}}
\put(50.00,14.00){\makebox(0,0)[rc]{$Y^{\varepsilon _1}_{J_1,n_1}$}}
\put(100.00,14.00){\makebox(0,0)[lc]{$Y^{\varepsilon _N}_{J_N,n_N}$}}
\end{picture}
\caption{The definition of the variables $s(k_i, e_i)$}
\end{center}
\end{figure}

To analyze the pole structure of $A_{N}$ it is
convenient to use the variables (see Fig. 1)
\begin{equation}
s (k_i,e_i)=k_i^2 -(e_i - \sqrt{2})^2 + 2
\label{s}
\end{equation}
associated with i-th internal line of the graph. Here $k_i$ and $e_i$ are
the momentum and Liouville energy for the line. Short comment is in order.
The unusual term in the definition of $s_i$ has it origin in the linear
term proportional to $Q$ in the operator $L_0$:
\begin{equation}
L_0=\frac{1}{2}[p_x^2-(p_{\phi} -\sqrt{2})^2+
2{\hat N}]
\label{l}
\end{equation}
where
\begin{equation}
{\hat N}= \sum_{n=1}^{\infty}(\alpha _{-n}\alpha _{n} + \phi _{-n}\phi _{n}
+n c_{-n}b_{n} +nb_{-n}c_{n})
\label{N}
\end{equation}
is the level number operator. It is easy to check that $s_i$
defined by eq.(\ref{s}) does not depend on what energy-momenta are taken
to define it: from the "left" or from the "right" of the $i$-th line.
When all the external states are on-shell then
\begin{equation}
s_i=s_i^0 =2[(\sum n_r )^2 -(\sum (1- \sigma_r J_r))^2
+2\sum (1-\sigma_r J_r)]
\label{s0}
\end{equation}
and with this notation the poles of a graph will appear as $1/(s_i-s_i^0)
=1/\delta s_i$.  For each $N$-point tree graph the number of variables
$s_i(k,e)$ is equal to $N-3$ while the number of independent variables
$\varepsilon _r$ is $N-1$ ($N$ external legs minus one conservation low).
In the following it will be convenient for us to parameterize each
$N$-point amplitude by $s_i$. To have a one to one correspondence between
$\{s_i\}$ and $\{ \varepsilon _r \}$ we fix a pair of $\varepsilon _r $
for two distinct vertices each having two external legs.
By using the definition (\ref{s}) we can obtain the relation between
$\delta s_i$ and $\delta \varepsilon _r =\varepsilon _r -\sqrt{2}(1-\sigma
_r J_r)$ up to terms of order $(\delta \varepsilon _r)^2$
$$
s_i-s^0_i=-2(e_i - \sqrt{2})\delta e_i + O((\delta e_i)^2),
$$
\begin{equation}
e_i=\sum \varepsilon _r,~~~\delta e_i=\sum \delta\varepsilon _r,
\label{con}
\end{equation}
where the sum in the second line of (\ref{con}) is performed over all external
lines standing from the left (or from the right) of $i$-th internal line.

In the following we shall examine only the graphs such that the
corresponding amplitudes have poles in all the variables $s_i$ which
parameterize the graph. We shall call graphs of this type
the {\em most singular graphs}. As it was shown in {\bf I} and
\cite{az2} there exist graphs which by dynamical reasons do not
have poles in some $s_i$.

As a preliminary step let us consider the calculation of the typical
element of a graph: a result of contraction of a pair of off-shell states
with the three string Witten vertex
$_1\langle \,_2\langle \,_3\langle V|$
\begin{equation}
_1\langle \,_2\langle \,_{3'}\langle V|
 Y_{a_1} \rangle _1
 |Y_{a_2} \rangle _2
|I\rangle _{3'}\rangle _{3}
\label{v}
\end{equation}
Here we have transformed the bras into kets by using the "identity "
$|I\rangle \rangle $. As it was shown in \cite{kaku} the
"identity" operator and three-string vertex are BRST-invariant
\begin{equation}
_1\langle \,_2\langle \,_3\langle V|
(Q^{(1)} +Q^{(2)} +Q^{(3)})
 =0,
\label{qv}
\end{equation}
\begin{equation}
(Q^{(1)} +Q^{(2)} ) |I\rangle_1\rangle_2 =0 ,
\label{qi}
\end{equation}
so for two on-shell states one has
\begin{equation}
Q^{(3)}\,
_1\langle \,_2\langle \,_{3'}\langle V|
 Y_{a_1} \rangle _1
 |Y_{a_2} \rangle _2
|I\rangle _{3'}\rangle _{3}=0.
\label{v0}
\end{equation}
Note, that energy conservation low in $|I\rangle \rangle $ and
$\langle \langle \langle V |$ in presence of the background
charge is modified to give: $\delta (\sum \varepsilon  -2\sqrt{2})$.
The BRST-cohomology for ghost number 2 was described in \cite{bou} that
results in:
\begin{equation}
\,_1\langle \,_2\langle \,_{3'}\langle V| Y_{a_1}
 \rangle _1 |Y_{a_2} \rangle _2 |I\rangle _{3'}\rangle _{3}= f_{a_1a_2}
 ^b c_0|Y_{b}\rangle _3+Q| \Lambda_{ a_1 a_2 }\rangle _3.
\label{v1}
\end{equation}
Now we take into account the explicit dependence of the vertex
$ _1\langle \,_2\langle \,_{3'}\langle V|$
on the Liouville energy.
Really, the relevant part of the vertex can be written as
$$
_1\langle \,_2\langle \,_{3'}\langle V|
\sim \int
d\varepsilon _1
d\varepsilon _2
d\varepsilon _3
\delta (\varepsilon _1+
\varepsilon _2  +
\varepsilon _3  -2\sqrt{2})
$$
\begin{equation}
_1\langle \varepsilon _1|\,
_2\langle \varepsilon _2|\,
_3\langle \varepsilon _3|
\exp [\sum_{r=1}^{3}N^{rr}_{00}
(\varepsilon _r^2-2\sqrt{2}\varepsilon _r)+
\sum _{r,s=1}^{3}\sum_{n=1}^{\infty}N^{rs}_{0n}
\varepsilon _r\phi ^{(s)}_n],
\label{ver}
\end{equation}
where $N^{rr}_{00}$ and
$N^{rs}_{0n}$ are some numbers (the Neumann coefficients).
We see that the vertex depends analytically on $\varepsilon_r$ (at least when
acting on finite states), hence it follows from eq.(\ref{v1}) and the
definition of the off-shell states that
\begin{equation}
_1\langle \,_2\langle \,_{3'}\langle V|
 Y_{J_1 ,n_1 , \sigma_1}^{\varepsilon_1} \rangle _1
 |Y_{J_2 ,n_2 , \sigma_2}^{\varepsilon_2} \rangle _2
|I\rangle _{3'}\rangle _{3}
-f_{123}|Y_{J_3 ,n_3 , \sigma_3}^{\varepsilon_1 +\varepsilon_2 }\rangle _3
-Q| \Lambda_{J_1 n_1 \sigma_1 ,J_2 n_2 \sigma_2 }
^{\varepsilon_1 +\varepsilon_2 }\rangle _3 =O
(\delta\varepsilon_1 ,\delta\varepsilon_2).
\label{v2}
\end{equation}

Next we note that the states $Y^{\varepsilon}_a $diagonalize  $L_0$.
For the state $|Y^{e}_{a}\rangle $ of given energy-momentum
$(k,e )$ $=(\sqrt{2}(n_1+n_2),~ \varepsilon _1+\varepsilon _2)$
we have according to definition (\ref{s}) and (\ref{l}):
$L_0 |Y^e_a\rangle=(\frac{1}{2}s(k,e)+\hat{N}-1)|Y^e_a\rangle$
hence for a discrete states $Y^\pm _{J,n}$ one gets:
$(\hat{N}-1)|Y^\pm_{J,n}\rangle
=-\frac{1}{2}s^0 (n,J)|Y^\pm_{J,n}\rangle $.
Taking into account our definition of the off-shell states we have
\begin{equation}
\label{ly}
L_0 |Y^{e}_{a}\rangle  =\frac{1}{2}(s(k,e)-s^0 )
|Y^e _{a}\rangle =\frac{1}{2}\delta s (k,e)|Y^e_{a}\rangle .
\end{equation}
Making use of eqs. (\ref{v2})  and (\ref{ly}) we see that the element
(\ref{v}) being multiplied by the propagator $\frac{b_0}{L_0}$ is
equal to
$$
_1\langle \,_2\langle \,_{3'}\langle V|
 Y_{a_1}^{\varepsilon_1} \rangle _1
 |Y_{a_2}^{\varepsilon_2} \rangle _2
\frac{b_0^{(3)}}{L_0^{(3)}}
|I\rangle _{3'}\rangle _3=
\frac{2}{\delta s(k,e)}f_{a_1 a_2}^{ b}|Y_b^e\rangle _3
$$
\begin{equation}
\label{blok} +
\frac{b_0^{(3)}}{L_0^{(3)}}
Q^{(3)}|\Lambda ^e_{a_1 a_2}\rangle _3 +
\frac{b_0^{(3)}}{L_0^{(3)}}
[\delta \varepsilon _1 |\Omega _{1 a_1 a_2}\rangle +
\delta \varepsilon _2 |\Omega _{2 a_1 a_2}\rangle ]
\end{equation}
with $k=\sqrt{2}(n_1+n_2)$ and $ e=\varepsilon _1 + \varepsilon _2$.
The substitution of eq.(\ref{blok}) into the full graph leads to a
representation
of the original graph as a sum of new graphs. Let us extract only the terms
in this sum which determine the most singular behavior of the original
graph.

The first term in eq.(\ref{blok}) contains a simple pole in variable $s$.
It diverges as soon as the external states 1 and 2 move on shell and therefore
will give  a contribution to the most singular part of the original graph.

The third term in eq.(\ref{blok}) does not lead to any infinities
in on-shell limit: a possible pole originating from $\frac{1}{L_0}$
will be cancelled by factors $\delta \varepsilon _1$
and $\delta \varepsilon _2$.
Therefore this term does not produce the most singular graphs.

The remaining second term in eq.(\ref{blok}) can be rewritten in the form
\begin{equation}
|\Lambda ^e_{a_1 a_2} \rangle-
Q\frac{1}{L_0}\rangle
|\Lambda ^e_{a_1 a_2}\rangle .
\label{lam}
\end{equation}
The first term in eq.(\ref{lam}) is finite in on-shell limit. The second
term contains the BRST operator $Q$ which acts on one of the vertices of
the remaining part of whole graph. Using the properties (\ref{qi}) and
(\ref{qv}) one can pull $Q$ through the graph until it acts on some
external line. As a result we obtain the new series of graphs with
cancelled propagator (owing to the identity $[Q,\frac{b_0}{L_0}]=1$) as
well as the series of graphs with operator $Q$ acting on one of the
external legs.  Each of these new graphs will have the lower singular
behavior as compared to the original graph.

So we see that the most singular behavior of original graph is
determined by
\begin{equation}
\label{hs}
A_N^{m.s.}(a_1,...,a_N)=\frac{2}{s_{b_1}-s^0_{b_1}}
f_{a_1 a_2}^{b_1}
A_{N-1}(b_1, a_3,...,a_N).
\end{equation}
In the r.h.s. of this expression we have the product of the pole,
the structure consistent and the $N-1$-point amplitude corresponding to
the graph in which the element (\ref{v}) is replaced by the discrete state
$Y^{\varepsilon _1+\varepsilon _2}_{b_1}$.

On the next step we can repeat the above procedure for $A_{N-1}$:
\begin{equation}
\label{hs1}
A_{N-1}^{m.s.}(b_1,a_3...,a_N)=\frac{2}{s_{b_2}-s^0_{b_2}}
f_{b_1 a_3}^{b_2}
A_{N-2}(b_2, a_4,...,a_N).
\end{equation}

Going step by step we come to the following expression for the most
singular part of $A_N$:
\begin{equation}
\label{ams}
A_N^{m.s.}(a_1,...,a_N)=(\prod _{i=1}^{N-3}
\frac{2}{s_{b_i}-s^0_{b_i}})
f_{a_1 a_2}^{b_1}
f_{b_1 a_3}^{b_2}...
f_{b_{N-3} a_{n-1}}^{a_{n}}
\end{equation}

\section{Traversal-like gauge and effective Lagrangian}
The aim of this section is to find a way to separate out the discrete states
interaction singularities just on the Lagrangian level by choosing
an appropriate gauge.

We put the following gauge condition on a string field $|\Phi
\rangle = |\varphi \rangle +c_0 |\psi \rangle$:
\begin{equation}
\tilde{Q}|\varphi\rangle
=0
\label{gauge}
\end{equation}
\begin{equation}
|\psi \rangle ~\mbox{-- arbitrary},
\end{equation}
where $\tilde{Q}$ is defined by the ghost zero modes expansion of $Q$
: $Q=c_0 L_0 +b_0 M +\tilde{Q}$.
Equation (\ref{gauge}) is well known in the literature as
defining the relative cohomology of $Q$ \cite{bou}. However, we do not
restrict the space of possible solutions for eq.(\ref{gauge}) to be
on-shell, i.e. we do not put the usual constraint: $L_0 |\varphi\rangle
=0 $, so $\tilde{Q}^2 =-L_0M\neq 0$.

The gauge $\tilde{Q} |\varphi\rangle =0$ is admissible. Indeed, the
gauge transformation $\delta |\Phi\rangle =Q|\Lambda\rangle$
with $|\Lambda \rangle =|\lambda \rangle +c_0|\omega \rangle$
for the field $|\varphi\rangle$ gives
\begin{equation}
\delta |\varphi\rangle =\tilde{Q}|\lambda\rangle +M|\omega\rangle,
\label{del}
\end{equation}
 Suppose $\tilde{Q}|\varphi\rangle \neq 0$ then
\begin{equation}
\tilde{Q}(|\varphi\rangle +\delta| \varphi\rangle )=\tilde{Q}
|\varphi\rangle -L_0 M|\lambda \rangle +\tilde{Q} M|\omega\rangle .
\label{q}
\end{equation}
Put $|\omega\rangle =0$, then the gauge condition will be satisfied
by $|\lambda\rangle =M^{-1}L_0^{-1}\tilde{Q} |\varphi\rangle$. The
crucial point here is if $M=\sum_{n} nc_{-n}c_n$ is invertible or not.
The resolution actually depends on the ghost number of a state
$|\lambda\rangle$. In our case $N_{gh}(|\lambda\rangle)=0$. We have
not proved rigorously that $M$ is invertible for this ghost number, but the
analysis of a number of examples forces us to conjecture that this
is actually the case. Hence, we conclude that our gauge condition
eq.(\ref{gauge}) can be fulfilled.

One important remark is now in order. It is obvious from the
discussion in the previous section that the most singular are the
graphs which join three discrete states in each three-string vertex. The
explicit form of the structure constants \cite{KP} tells us
that in this case the vertex has two $+$ legs and one $-$ leg and
the propagator connect one $+$ and one $-$ state. If we denote by $N_+$
($N_-$) the number of external $+$ ($-$) legs for a graph with $V$
vertices, $\nu$ loops and ${\cal N} =\nu + V-1$ internal lines,
then we get:
$V=N_- +{\cal N} =N_- +V-1+\nu$, i.e.,
\begin{equation}
\label{loop}
N_- =1-\nu .
\end{equation}
So, we recognize that just the tree graphs are of special
significance in the model and in the following we shall convince
ourselves by these graphs.

Thus, for our specific task we can ignore the ghost contribution
and overcome the complete gauge fixing procedure restricting
ourselves by the first step, i.e. by taking the partition function
in the form:
\begin{equation}
Z=\int [{\cal D}\varphi ][{\cal D}\psi ]\delta (\tilde {Q} \varphi ) e^{-S(
\varphi ,\psi)}
\label{z}
\end{equation}

Now let us discuss the actual content of the gauge fixing condition $\tilde{Q}
|\varphi\rangle =0$. The string field in momentum representation can
be written down in the form:
\begin{equation}
|\varphi\rangle =\int d\varepsilon \sum_{A,n}\varphi_{A,n} (\varepsilon )M_A (
\alpha , \phi ,b,c)|\sqrt{2}n,\varepsilon \rangle^{x,\phi
}|1\rangle^{b,c},
\label{fi}
\end{equation}
where $\varphi_{A,n}(\varepsilon )$ are the ordinary fields, $M_A$ --
some monomials in string ($\alpha_k ,\phi_k$)
and ghost normal modes, the ghost vacuum
$|1 \rangle ^{bc}$
is defined via $SL_2$ invariant one by
$|1 \rangle ^{bc}=
c_1|0 \rangle ^{bc}$ and we have
utilized the fact that the "space" is compactified on $SU(2)$
radius. The gauge fixing condition results in an infinite system of
linear equations that can be written symbolically as
\begin{equation}
\varepsilon C_{AB} \varphi_{Bn}(\varepsilon )
+D^n_{AB}\varphi_{Bn}(\varepsilon )=0.
 \label{si}
\end{equation}
For each
fixed values of $n$ and $\varepsilon $ and fixed mass level only a finite
number of equations in (\ref{si}) survives thus giving a possibility to
solve the system. So, we see that not all the integration variables in
$[\cal{D} \varphi ]$ are independent but there is a arbitrariness in
choosing the proper integration variables.

To exemplify our choice let us consider the mass level
one. The $|\varphi \rangle$ component of the string field
$|\Phi\rangle$ on this level is given by
\begin{equation}
|\varphi\rangle =\int d\varepsilon \sum_n [A_n (\varepsilon )\alpha_{-
1} +D_n
(\varepsilon )\phi_{-1} ] |\sqrt{2}n,\varepsilon \rangle^{x,\phi}|1\rangle^
{b,c}
\label{1a}
\end{equation}
The gauge condition results in single linear equation:
\begin{equation}
\sqrt{2} n A_n (\varepsilon ) -(\varepsilon -2\sqrt{2})D_n
(\varepsilon )=0.
\label{2a}
\end{equation}
For $n=0$ the field $A_0 (\varepsilon )$ remains to be an arbitrary
function of $\varepsilon$ while the value of $\varepsilon$ for $D_0
(\varepsilon )$ is fixed to be: $\varepsilon =2\sqrt{2}$ that is nothing
but the value defined by the mass-shell condition for this level:
\begin{equation}
2n^2 -\varepsilon (\varepsilon -2\sqrt{2} )=0.
\label{ml}
\end{equation}
On the other hand, the state $\alpha_{-1} |0,\varepsilon\rangle^{x,
\phi}|1\rangle^{b,c}$ corresponds to the off-shell discrete states
$|Y^{\varepsilon} _{1,0}\rangle$, while \begin{equation}
\phi_{-1}|0,2\sqrt{2}\rangle^{x,\phi} |1\rangle^{b,c}\equiv
L_{-1}|0,2\sqrt{2}\rangle^{x,\phi} |1\rangle^{b,c} =
\tilde{Q} |0,2\sqrt{2} \rangle^{x,\phi}|0\rangle^{b,c}
\label{3a}
\end{equation}
is one of the spurious states at discrete values of momenta
that have been studied in \cite{kor}. The complete contribution to
the string field $|\varphi\rangle$ from $n=0$ fields reads:
\begin{equation}
|\varphi^0 \rangle =\int d\varepsilon A_0 (\varepsilon )|Y^\varepsilon _{1,0}
\rangle +
\frac{1}{\sqrt{2}}D_0 (2\sqrt{2}) \tilde{Q}|0,2\sqrt{2}\rangle^{x,\phi
}|0\rangle^{b,c}.
\label{n0}
\end{equation}
For $n\neq 0$ eq.(\ref{2a}) yields:
\begin{equation}
A_n (\varepsilon ) =\frac{-1}{\sqrt{2}n} (\varepsilon -2\sqrt{2} )D_n
(\varepsilon).
\label{4a}
\end{equation}
After the suitable redefinition of the field $D_n$ one gets the
contribution into $|\varphi\rangle$:
\begin{equation}
|\varphi^1 \rangle =\int d\varepsilon \sum_{n\neq 0}D_n (\varepsilon)
[\frac{2L_0}{\sqrt{2}
n} \alpha_{-1}c_1 -
2\tilde{Q}]|\sqrt{2}n,\varepsilon \rangle^{x,\phi}|0\rangle^{bc}
\label{5a}
\end{equation}
As it must be, the decomposition of the string field being restricted
on mass-shell strictly reproduces the well known results on
relative cogomology \cite{bou}.

The lesson that one learns from this example is the following,
any string field $|\varphi\rangle$ obeying the gauge condition
(\ref{gauge}) can be presented in the form:
\begin{equation}
|\varphi\rangle = \int d\varepsilon [A_{J,n}(\varepsilon )|
Y^{\varepsilon}_{J,n}\rangle +D_{a,n}(\varepsilon)|a,\sqrt{2}n,\varepsilon
\rangle],
\label{phi}
\end{equation}
where
$a$ is a label of states and each
 state $|a,\sqrt{2}n,\varepsilon\rangle$ being restricted
on-shell is $\tilde{Q}$ exact. This decomposition provides the proper
integration variables in this gauge as $A(\varepsilon )$ and
$D(\varepsilon)$.

In the following we shall use for the decomposition (\ref{phi})
the shorthand
\begin{equation}
|\varphi\rangle =|A\rangle +|D\rangle .
\label{ph}
\end{equation}
With this notation the gauge fixed quadratic action can be written as
\begin{eqnarray}
S_0 & = & \int A \ast c_0 L_0 A +2\int A
\ast c_0 L_0 D \\
& & +\int D \ast c_0 L_0 D +\int \psi \ast
c_0 M\psi
\label{ss0}
\end{eqnarray}
The action defines three types of internal lines and it is obvious
that only $\langle A,A\rangle$ and $\langle D,D\rangle$ lines
can be relevant for the most singular graphs.

Let us define the projection operator:
\begin{equation}
P(\varepsilon)|\varphi(\varepsilon )\rangle =|D(\varepsilon )\rangle
\label{pro}
\end{equation}
which can be expanded over the eighenspaces of the mass level
operator $\hat{N}$
\begin{equation}
P=\sum_{N=1}^{\infty}\sum_{n=-\infty}^{\infty}P_{N,n}(\varepsilon)
,
\label{pn}
\end{equation}
where $N$ is the eighenvalue of $\hat{N}$. For $\varepsilon$
on-shell
\begin{equation}
2n^2 -(\varepsilon -\sqrt{2})^2 +2N=0,
\label{ep}
\end{equation}
i.e. for
\begin{equation}
\varepsilon =\varepsilon^\sigma_{N,n}=\sqrt{2}(1-\sigma\sqrt{n^2
+N}),~~\sigma =\pm ,
\label{eps}
\end{equation}
one has due to eq.(\ref{pro})
\begin{equation}
P_{N,n}(\varepsilon^\sigma _{N,n})=\tilde{Q}\times R^\sigma _{N,n}
=Q\cdot R^\sigma _{N,n}
\label{pe}
\end{equation}
for some operators $R^\sigma _{N,n}$.

Let us examine the singular structure for $\langle D,D\rangle$
lines. Writing down the projector (\ref{pn}) explicitly one has
for the $D$-fields propagator:
\begin{equation}
P^+ \frac{b_0}{L_0}P=\sum_{N,n}P^+_{N,n}(e)
\frac{2b_0}{2n^2 -(e-\sqrt{2})^2 +2N}P_{N,n}(e).
\label{pr}
\end{equation}
For $e$ in some neighborhood of a spectral point $e_0$
$ 2n_0^2 -(e_0 -\sqrt{2})^2 +2N_0 =0$ one has for the numerator owing to
eq.(\ref{pe}):
\begin{eqnarray}
P^+_{N_0 ,n_0}(e)b_0 P_{N_{0},n_0}(e)&=&[QR_{N_0 ,n_0}(e_0 ) +O(e-e_0
)]^+ b_0 [QR_{N_0 ,n_0}(e_0 )+O(e-e_0 )] \\ \nonumber
&=& R^+_{N_0 ,n_0} Qb_0 QR_{N_0 ,n_0} +O(e-e_0 )\\ \nonumber
&=& R^+_{N_0 ,n_0} L_0 R_{N_0 ,n_0} +O(e-e_0 ).
\label{pr1}
\end{eqnarray}
Thus, we see the cancelation of the potential pole.

Hence, we conclude that only the trees build from $\langle A,A
\rangle$ lines can produce the desired singularities. The quadratic
part of the action for $A$ fields reads (see eq.(\ref{phi}))
\begin{equation}
S_0 (A)=\frac{1}{2}\int d\varepsilon \sum_{J=0}^{\infty}\sum_{n=-J}
^J A_{J,n}(-\varepsilon +2\sqrt{2})[(\varepsilon
-\sqrt{2})^2 -2J^2 )]A_{J,n}(\varepsilon)
\label{so}
\end{equation}
To obtain the interaction term $S_{int}$ recall that it is defined
by the correlation function
on Witten string configuration $R_W$:
\begin{equation}
\,_1\langle \,_2\langle \,_{3}\langle V|
Y_{J_1 ,n_1 , \sigma_1}^{\varepsilon_1} \rangle _1
|Y_{J_2 ,n_2 , \sigma_2}^{\varepsilon_2} \rangle _2
|Y_{J_3 ,n_3 , \sigma_3}^{\varepsilon_3}\rangle _{3}
=\langle
Y^{\varepsilon_1}_{J_1 ,n_1}(w_1)
Y^{\varepsilon_2}_{J_2 ,n_2}(w_2)
Y^{\varepsilon_3}_{J_3 ,n_3}(w_3) \rangle _{R_W}.
\label{int}
\end{equation}

After a suitable conformal
transformation to upper half-plain it is calculated to give:
$$
S_{int}(A)=\frac{2}{3}\int d\varepsilon_1 d\varepsilon_2
d\varepsilon_3
\delta (\varepsilon_1 +\varepsilon_2 +\varepsilon_3 -2\sqrt{2})$$
\begin{equation}
\cdot\sum_{J_1 n_1}
\sum_{J_2 n_2}
\sum_{J_3 n_3}
F_{J_1 ,n_1 ;J_2 ,n_2 ;J_3 ,n_3}(\varepsilon_1 ,\varepsilon_2 ,\varepsilon_3)
A_{J_1 n_1}(\varepsilon_1)
A_{J_2 n_2}(\varepsilon_2)
A_{J_3 n_3}(\varepsilon_3)
,
\label{int1}
\end{equation}
where
\begin{equation}
F_{J_1 ,n_1 ;J_2 ,n_2 ;J_3 ,n_3}(\varepsilon_1 ,\varepsilon_2 ,\varepsilon_3)
= \lim_{x\to \infty }x^{2(\varepsilon_1 -\sqrt{2})^2 -4J_1^2}
\prod_{r=1}^3 (\frac{4}{3\sqrt{3}})^{(-\varepsilon_2 -\sqrt{2})^2 +2
J_2^2}
\langle
Y^{\varepsilon_1}_{J_1 ,n_1}(x)
Y^{\varepsilon_2}_{J_2 ,n_2}(1)
Y^{\varepsilon_3}_{J_3 ,n_3}(0) \rangle .
\label{int2}
\end{equation}

It is not difficult to realize that the action $S=S_0 +S_{int}$
strictly reproduces the formula (\ref{ams}) for the most singular
part by following the usual Feynman rules.
%%%%%%%%%%%%%%%%%%%%%%%%%%%%%%%%%%%%%%%%%%%%%%%%%%%%%%%%%%%%%%%%
\section{Conclusion}
Let us make some general comments on the problem
of constructing the tree-level $S$-matrix for 2d string discussed in this
paper.

Our system has common features with the usual quantum field theory in the
finite
volume. Indeed, in Quantum Field Theory in the finite volume one deals with
discrete
states with momenta lying on a lattice. The  $S$-matrix approach for such
a theory is not adequate since the free Hamiltonian has the discrete
spectrum and as it is well known from Quantum Mechanics, the $S$-matrix
approach is not a proper one for systems with a discrete spectrum. In our
case the situation is more complicated since we have to deal with the
 system  resembling a field theory model with particle decays.
These facts have to push us to consider the off-shell $S$-matrix
approach and String Field Theory as a starting point.

In the paper we restrict ourself
with consideration of
tree-level scattering amplitudes. This is related with the fact that
according to relation (\ref{loop}) the most singular parts
of the off-shell $S$-matrix are concentrated in the non-loop diagramms.

Note also that there is an alternative possibility to avoid the
appearance of singularities at all. It consists in introduction of some
extra factors for vertex operators.
As it has been shown above the singular part of
each tree graph is the product of poles and the structure constants, which
are subject of OPE of external states vertex operators.  To get total
amplitudes one has to make all permutations of external states.  Since all
of the structure constant (\ref{ff}) entering in OPE of vertex operators
with cocycle factors are antisymmetric in low indices, it is follows that
the most singular parts of the graphs will be cancelled in the total
scattering amplitudes. So in the case when the external states are
represented by vertex operators with cocycle factors, the decay situation
is not admitted.  However behind this construction
we have not selfconsistent scheme such as String Field Theory.
\vspace{10mm}

{\bf Acknowledgment}

This work is supported in part by RFFR under grant N93-011-147 and ISF
under grant M1L-000.

%%%%%%%%%%%%%%%%%%%%%%%%%%%%%%%%%%%%%%%%%%%%%%%%%%%%%%%%%%%%%%%%%

\end{document}